\documentclass[%
 reprint,
 amsmath,amssymb,
 aps,
]{revtex4-2}

\usepackage{graphicx}
\usepackage{subfigure}
\usepackage{dcolumn}
\usepackage{bm}
\usepackage{enumitem}

\begin{document}

\preprint{APS/123-QED}

\title{Statistical Properties of Links of Network: A Survey on the Shipping Lines of Worldwide Marine Transport Network}
\author{Wenjun Zhang}

\author{Weibing Deng}

\email{wdeng@mail.ccnu.edu.cn}
\author{Wei Li}

\email{liw@mail.ccnu.edu.cn}
\address{Complexity Science Center, Key Laboratory of Quark and Lepton Physics (MOE), Central China Normal University, Wuhan 430079, China}

\date{\today}

\begin{abstract}
Node properties and node importance identification of networks have been vastly studied in the last decades. While in this work, we analyse the links' properties of networks by taking the Worldwide Marine Transport Network (WMTN) as an example, i.e., statistical properties of the shipping lines of WMTN have been investigated in various aspects: Firstly, we study the feature of loops in the shipping lines by defining the line saturability. It is found that the line saturability decays exponentially with the increase of line length. Secondly, to detect the geographical community structure of shipping lines, the Label Propagation Algorithm with compression of Flow (LPAF) and Multi-Dimensional Scaling (MDS) method are employed, which show rather consistent communities. Lastly, to analyse the redundancy property of shipping lines of different marine companies, the multilayer networks were constructed by aggregating the shipping lines of different marine companies. It is observed that the topological quantities, such as average degree, average clustering coefficient, etc., increase smoothly when marine companies are randomly merged (randomly choose two marine companies, then merge the shipping lines of them together), while the relative entropy decreases when the merging sequence is determined by the Jensen-Shannon distance (choose two marine companies when the J-S distance between them is the lowest). This indicates the low redundancy of shipping lines among different marine companies.

\end{abstract}

\maketitle


\section{Introduction}
Marine transportation account for more than 70 per cent of global trade \cite{unctad}. For many major economies, a even larger proportion of their import and export activities may rely heavily on marine transportation (e.g. about 90\% as for China). The advantages of the marine transport \cite{haezendonck2000competitive} include high throughput, low expense and less restricted access to the shipping lines. However, the rapid development of world marine transport brought a series of challenges as well, such as, how to raise the transport ability? How to arrange the shipping lines of different marine companies to avoid cutthroat competition? All of these issues are associated with the links' properties of WMTN. With the development of complex network theory, a wide range of traffic networks have been studied in the framework of complex networks, such as China air transportation \cite{Li2004Statistical}, Urban Traffic \cite{Hu2006Urban}, Europe railway
\cite{Puffert2002Path,marti2013european}, an world marine transport \cite{wei2009worldwide}.

The previous studies on World Marine Transport Network (WMTN) mostly focused on the topological properties. Deng et al. \cite{wei2009worldwide} studied topological structure of World Marine Transport Network (WMTN), such as degree distribution, clustering coefficient, average shortest-path length, efficiency, etc. Kaluza et al. \cite{kaluza2010complex} compared the topological properties of WMTN with those of the sub-networks of container transportation, dry goods transportation and oil transportation, respectively. Such as number of lines, throughput, average degree, cluster coefficient, average distance, power law distribution index, community, etc. They also studied the weight distribution and seaport distance distribution of WMTN. Ducruet Csar and Notteboom Theo \cite{ducruet2012worldwide} studied the evolution of WMTN which contains the throughput of WMTN from 1972 to 2008, seaports number, edges length, loops number, efficiency, power distribution index, and assortativity, etc. In addition, the relevant visualizations were provided.

While the node properties and node importance identification of WMTN have been studied extensively \cite{wei2009worldwide,kaluza2010complex,ducruet2012worldwide}, investigations on the links' properties are lacking. In this work, we explored the statistical features of the shipping lines in the WMTN. Firstly, we studied the four quantities of nodes centrality which are degree, betweenness, seaport weight and seaport throughput, so as to show the importance of different seaports. Then we defined the quantity of line saturability to measure lines' transport ability. The feature of the community structure in the WMTN is important for the arrangement of shipping lines but still largely unknown at this moment. We divided the WMTN into seven communities by Label Propagation Algorithm with compression of Flow (LPAf) \cite{Han2016Community}, which can be recognized as one of the state of the art algorithm in community detection. The detected community results are then compared to the seaports' geographical distribution and multi-dimensional scaling (MDS) \cite{borg2005modern}. In the procedure of MDS, we introduced the diffusion distance \cite{segarra2015diffusion} to be the seaports distance. The community structure is well defined if the seaports of the same community are classified into one cluster. Lastly, we studied the relation between different marine companies by means of multilayer networks. Each marine transport company is represented by a single layer of the network, with aggregation taking place randomly from numbers 1 to 42. Each aggregation yields a network, by which we calculated edges number, mean degree, network efficiency, and average trapping time. Then we aggregated the marine companies according to the Jensen-Shannon distance \cite{Lin1991Divergence,Domenico2016Spectral,De2015Structural}, that is, if the Jensen-Shannon distance of two marine companies is the lowest, we merge the shipping lines of them together, and calculated the relative entropy \cite{De2015Structural,Feynman1963118}. We can know the relation between different marine companies by observing the change of above quantities with the number of marine companies aggregated.

The rest of this paper is organized as follows. In Section 2, we show some topological properties of WMTN and propose the definition of line saturability. In Section 3, the community structure was given. In Section 4, we introduce the multilayer networks and discuss the relation between marine companies. Finally, we conclude in Section 5.
\section{The topological properties of WMTN}
\label{sec:1}
The data used in this paper was collected from 42 world renowned marine companies, such as COSCO, ANL, APL, HNM, etc.\cite{COSCO2014,ANL2014,APL2014,HNM2014}, totally 632 seaports and 2283 shipping lines. Currently, the public transport networks are presented in two different topological representations, one of them is $L$ space \cite{sen2003small}, consisting of nodes representing the seaports and edges between any two consecutive stops along the shipping lines. The node degree $k$ is just the number of different shipping lines one needs to take from a given seaport, while the shortest path length between two arbitrary seaports is the minimal number of stops one needs to make, then the $L$ space is also named the space of stops. We construct WMTN in the $L$ space, and calculate its node number $N$, average degree $\langle K\rangle$, clustering coefficient $C$, average shortest-path length $\langle d\rangle$, radius $R$, network efficiency \cite{Latora2001Efficient,achard2007efficiency} $E$, and average line length (the length of line path) $\langle L_p\rangle$ (see Table~\ref{tab:topological}).
\begin{table*}[tbp]
\caption{Topological properties of WMTN. $N$ is the number of ports, $n$ is the number of edges, $\langle K\rangle$ is the average degree, $L_{est}$ is the longest line length, $C$ is the clustering coefficient, $\langle d\rangle$ is the average shortest-path distance, $R$ is radius of networks, $E$ is the networks efficiency, and $\langle L_p\rangle$ is the average line length (the length of line path).}
\label{tab:topological}
\centering
\begin{tabular}{lcccccccccc}
\hline\hline
Topological quantities &$N$&$\langle K\rangle$&$C$&$\langle d\rangle$&$R$&$E$&$\langle L_p\rangle$&\\ \hline
WMTN &632&12.7&0.388&3.86&6&0.289&9.38\\
Poland urban bus&152$\sim$2811&2.48$\sim$3.03&0.055$\sim$0.161&6.83$\sim$21.52&--&--&--\\
WATN &188&15.06&0.729&2.139&3&0.441&--\\
\hline\hline
\end{tabular}
\end{table*}

These topological properties are consistent with previous studies \cite{wei2009worldwide} in WMTN, which verifies the reliability of our data, and also indicates that the WMTN does not change much over the recent years. For study the characteristic of WMTN, we compare the WMTN with other kind of transport networks which are Poland urban bus networks \cite{sienkiewicz2005statistical} and World Air Transport Network (WATN). The Poland urban bus network have 22 networks that belong to 22 different cites in Poland, so its' data reliability is high and we can compare WMTN with all 22 urban bus networks. The node numbers of these networks range from 152 to 2811, and as transportation networks, these networks also function very similarly with marine transport. The WATN data was collected from 22 top Civil Air Transport and it contains 188 ports and 1416 edges. We found that the average degree of WMTN is larger than that of Poland urban bus network, but lesser than WATN a little. This is because the seaports of WMTN can be connected with both its nearby seaports and the long-range ones, and the ports of air can be connected with other ports in long distance directly. But the bus stations of Poland are rarely connected over the long distance. The clustering coefficient is the represent the closeness of neighbors of one node, the more high clustering coefficient mean the more connection between neighbors of one node. We can measure this quantity by calculate the density of triangles in networks. This quantity of WMTN is slightly larger than Poland urban, and lesser than WATN. This result maybe due to the short length of the lines in the WMTN. Short lines mean that there are a lot of triangle lines in WMTN. Shortest-path length on networks is the path with shortest length of two nodes in a network. And for the WATN, its average degree is larger than WMTN slightly, but its nodes number is much lesser than WMTN. It is mean that the WATN have more high edge occupation radio. The average shortest-path length of WMTN is smaller than every networks of Poland urban bus. Consider the number of WMTN nodes is among the numbers of different Poland urban bus networks. So WMTN have small-world effect apparently \citep{Watts1998Collective}, so same as WATN. The networks efficiency is defined as a average value of inverse distances between network nodes. The WMTN efficiency less than WATN but large than Boston subway \cite{Latora2001Efficient} which networks efficiency is 0.1 and the number of its nodes is 124 (The networks efficiency of Poland subway were not calculated in ref.\cite{sienkiewicz2005statistical}). Since the WATN have so short average shortest-path, we think it is not sense to compare the efficiency of WMTN with WATN. So we deem that WMTN have high efficiency, and mean WMTN have high robustness \cite{Berche2009Resilience}.

\begin{figure}[htbp]
\centering
\includegraphics[width=0.45\textwidth]{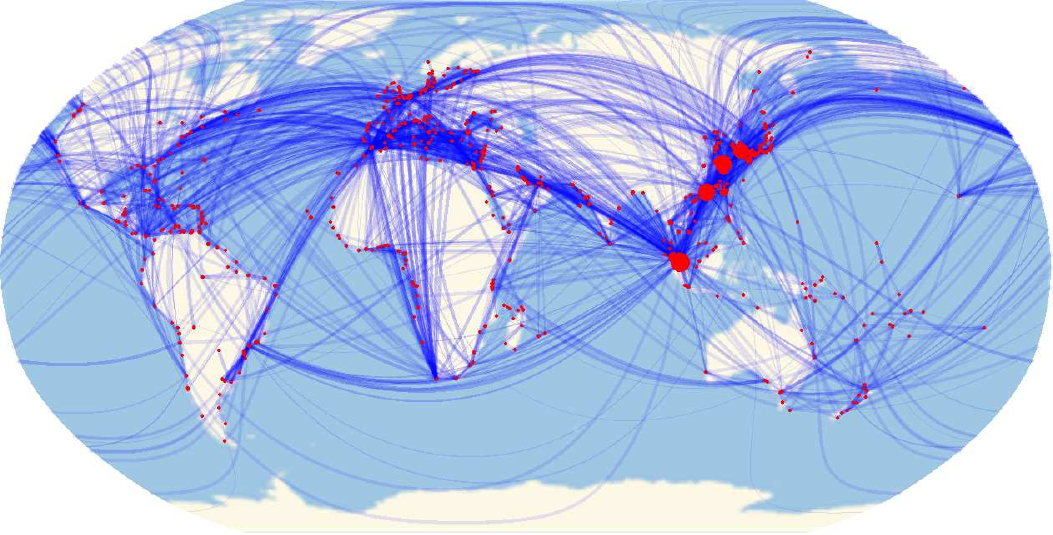}
\caption{(Color online) The geographical representation of the marine network, with the line thickness specifying its weight and the node size specifying the weight of the seaport.}
\label{fig:portline}
\end{figure}

We now turn to study the the importance of seaports which corresponding to nodes centrality, because such information is very useful for the organization of the marine lines. We found that some lines and seaports are very important, as shown in Fig.~\ref{fig:portline}. The degree centrality $K$ which is the degree of nodes and betweenness centrality $B$ is the number of shortest paths that pass through the node. But considering the specificity of WMTN, we introduce other quantities: weight (strength) $W$ and throughput $T$, to represent the importance of seaports. The weight $W$ is the number of shipping lines passing through this seaport and the throughput is the seaports transaction capacity in 2014 from data. Then we study the Pearson correlation coefficients between $K$, $B$, $W$, and $T$, which are given in Table.~\ref{Tab:pearson}. One can find there exists the strong Pearson correlation between $K$ and $B$ of the WMTN as well as W and T in the WMTN. This is because degree and betweenness are topological quantities of centrality of the network, while seaports weight and seaports throughput are actual transport quantities from metadata.

\begin{figure}[htbp]
\centering
\subfigure[A backtrack line with 5 seaports.]{
    \label{fig:subfig:a}
   \includegraphics[width=0.4\textwidth]{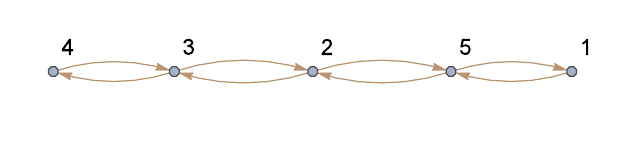}}
  \hspace{0.5in}
  \subfigure[A loop line with 5 seaports.]{
    \label{fig:subfig:b}
   \includegraphics[width=0.4\textwidth]{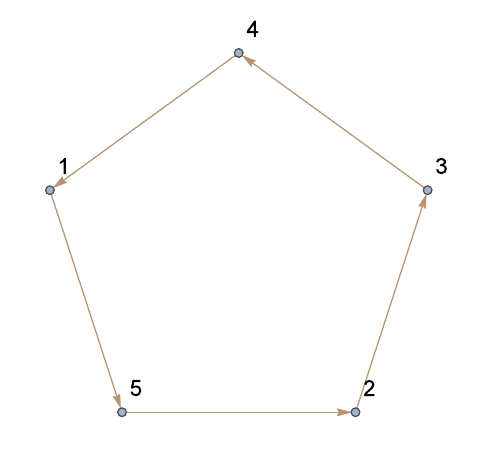}}
  \caption{(Color online) (a) The saturability is 0.625, where 8 steps are taken to provide transportation for 5 seaports. (b) The saturability is 1, where 5 steps are taken to provide transportation for 5 seaports.}
  \label{fig:linesaturability}
\end{figure}

\begin{table}[tbp]
\caption{The Pearson correlation coefficients between betweenness $B$, degree $K$, weight $W$, and throughput $T$ of WMTN, in a pairwise way. All the p-values are much less than 0.01.}
\label{Tab:pearson}
\centering
\begin{tabular}{llllll}
\hline\noalign{\smallskip}
Properties&$B$&$K$&$W$&$T$\\
\noalign{\smallskip}\hline\noalign{\smallskip}
$B$&1&{\bf{0.928}}&0.708&0.558\\
$K$&{\bf{0.928}}&1&0.805&0.664\\
$W$&0.708&0.805&1&{\bf{0.878}}\\
$T$&0.5581&0.664&{\bf{0.878}}&1\\
\noalign{\smallskip}\hline
\end{tabular}
\end{table}
As all lines of WMTN are loops, we can define line saturability by the loop property to quantify the operation ability of the shipping line. For a line with path length $L$ in WMTN and seaports number $N_p$, its lines saturability $S$ is
\begin{equation}
S=\frac{N_p}{L}
\end{equation}
Since one line never passes through one seaport three times, the value of line saturability is $0.5<S \le 1$, as shown in Fig.~\ref{fig:linesaturability}. Larger line saturability indicates higher operation ability of the shipping lines, because for same-length lines the more seaports are connected to it the more seaports can be reached by this line, so the more trade between different ports can be offered by this line.

\begin{figure}[htbp]
\centering
\includegraphics[width=0.45\textwidth]{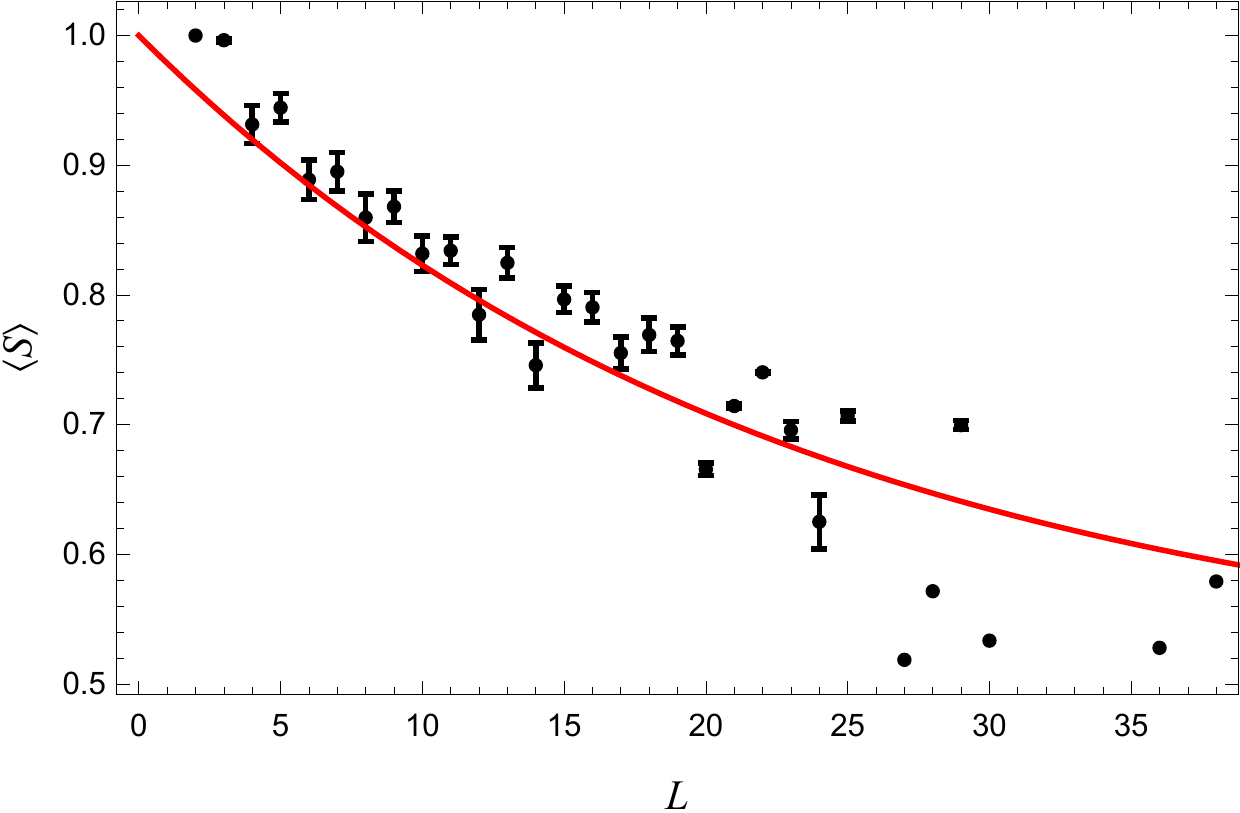}
\caption{(Color online) The average lines saturability $\langle S\rangle$ versus lines length $L$. The red line is the fitted line with the form $\langle S\rangle\propto 0.5+0.5 e^{-0.0437L}$ with standard error 0.0025 and t-statistic -17.184.}
\label{fig:saturability}
\end{figure}

The average line saturability of the WMTN is 0.877. It is a high value to compare with subway witch line saturability is 0.542. In view of the range of saturability, we set the fitting function of saturability is
\begin{equation}
\langle S\rangle\propto 0.5+0.5e^{a L}
\end{equation}
where $a$ is the parameter. We got the relation between line saturability and line length is shown in Fig.~\ref{fig:saturability} $\langle S\rangle\propto 0.5+0.5e^{-0.0437 L}$, the line saturability decreases exponentially with the line length. This feature might be due to the fact that the long shipping lines are usually set along the coast, so a ship along this line may pass through the seaports one more time when it
returns.
\section{Community structure of WMTN}
\begin{figure}[htbp]
\centering
\includegraphics[width=0.45\textwidth]{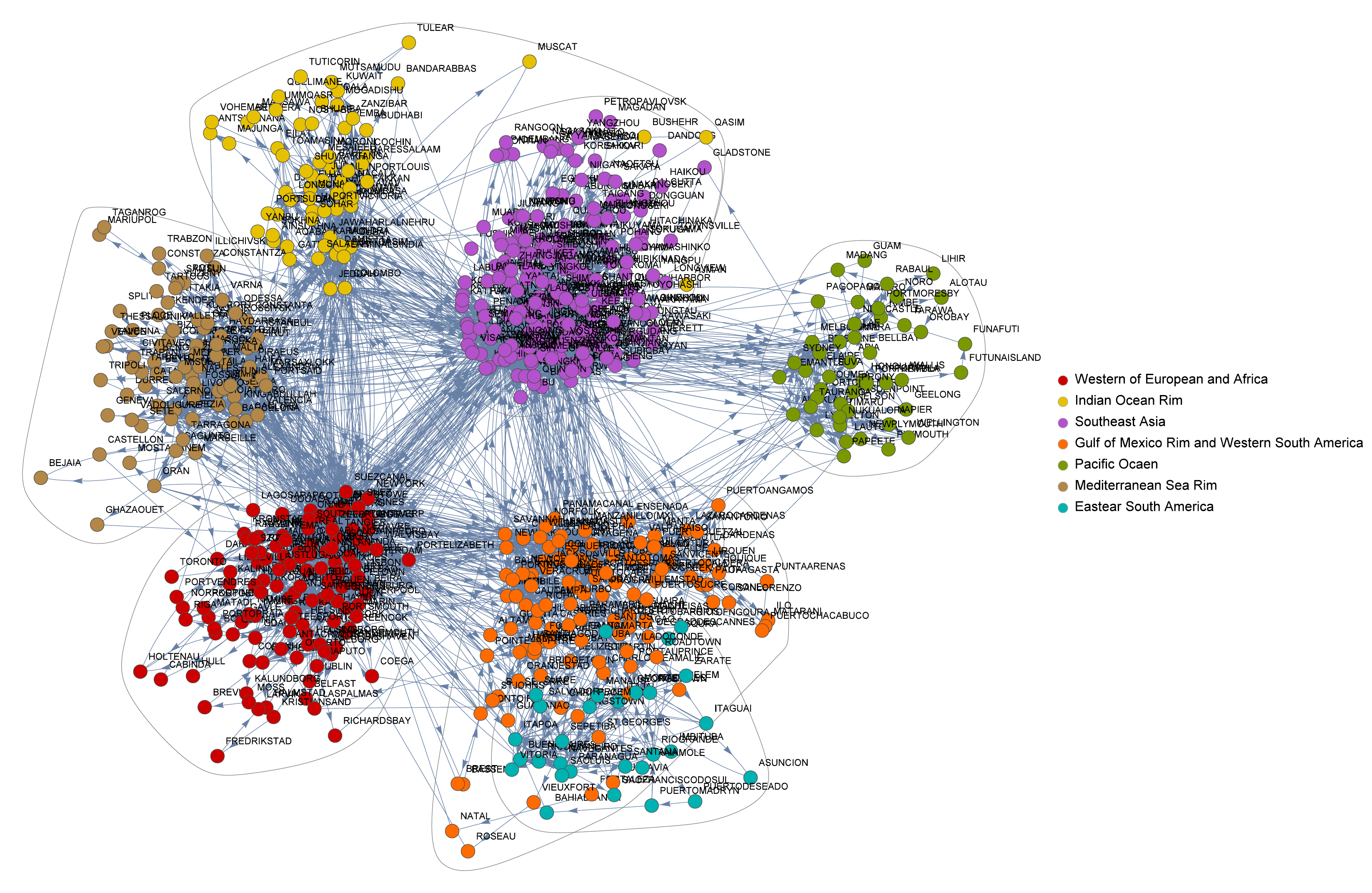}
\caption{(Color online) The community structure of WMTN. Different colors specify different communities and there are seven communities in total.}
\label{fig:community}
\end{figure}

\begin{figure}[htbp]
\centering
\includegraphics[width=0.5\textwidth]{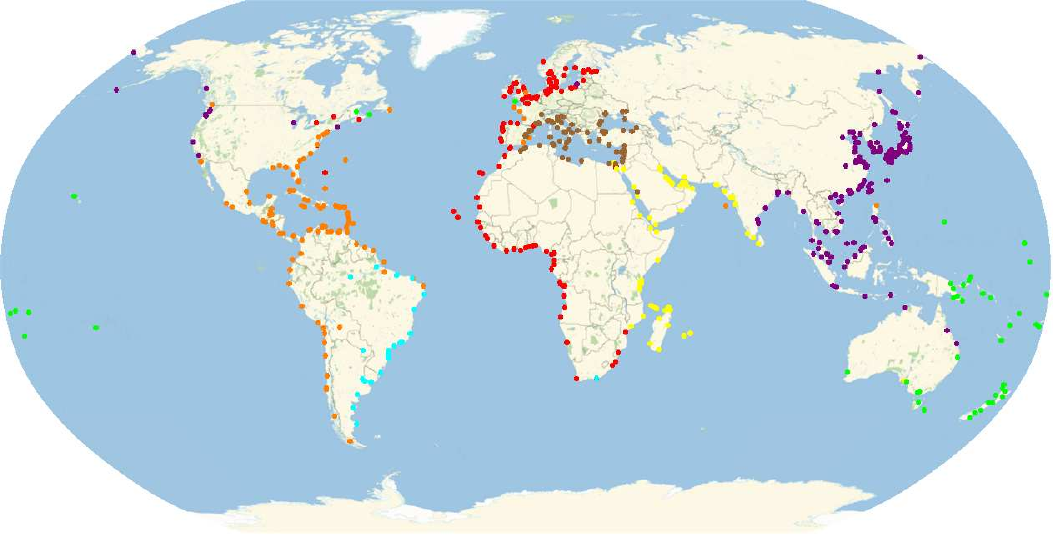}
\caption{(Color online) The seaports distribution on map. The colors' representation is the same community as in Fig.~\ref{fig:community}. It is obvious that one community tends to span a certain connected area and different communities are often separated by channels.}
\label{fig:portmap}
\end{figure}

The community structure of networks is a basic concept to study the structure of networks. It means that the network can be divide in some parts, and every parts contain more connection between the nodes in same part. M. Girvan and M. E. J. Newman \cite{girvan2002community} proposed the community concept and corresponding algorithm, such as Girvan-Newman algorithm, minimum-cut algorithm \cite{newman2004detecting}. The most widely used algorithm in this regard is modularity maximization \cite{newman2004fast}, proposed by Newman in 2004. The modularity also provides a criterion of justifying the goodness of community structure. But the community structure of WMTN was rarely studied in the past, except that Kaluza et al tried for the first time in \cite{kaluza2010complex}. Here we adopt the Label Propagation Algorithm with compression of Flow (LPAF) \cite{Han2016Community} to detect the community of WMTN. WMTN can be divided into seven communities (shown in Fig.~\ref{fig:community}) by using LPAF, and the modularity of the community structure is 0.59, suggesting that the LPAF is better than modularity maximization algorithm which yields 0.549 for the WMTN. Comparing the community structure to the seaports distribution on the geography map (represented by Fig.~\ref{fig:portmap}), one can see that the seaports of one community of WMTN are distributed closed on the geographical map. Hereby, the communities are as follows: Western Europe and Africa, Indian Ocean Rim, Southeast Asia, Gulf of Mexico Rim and Southern West America, Pacific Ocean, Mediterranean Sea Rim, and Southern East America. With this classification, one can find that the WMTN is quite geographically localized. So the radio of long distance transport is less than the radio of short distance transport, and the WMTN is limited by the geographical constraints, such as channels, coastlines, etc.

The goodness of community detection results is usually judged by modularity. Here we adopt the so-called Multi-Dimensional Scaling (MDS) to test the community structure through visualization. As we know, the MDS is a method used to visualize the level of similarity of individual cases of a dataset. For a set of individuals, each object is assigned a coordinate to make sure the distance between different individuals represents their similarity. Therefore the first step is to define the distance to quantify the similarity of individuals. And, for a set $X$ with a function $X\times X\to R$, and $x,y,z\in X$, we have 4 rules in metric space:
\begin{enumerate}
\item 
Non-negativity:
\begin{equation}
d(x,y)\geq 0
\end{equation}
\item Identity of indiscernibles
\begin{equation}
d(x,y)=0 \Leftrightarrow x=y
\end{equation}
\item
Symmetry
\begin{equation}
d(x,y)=d(y,x)
\end{equation}
\item
Subadditivity or triangle inequality
\begin{equation}
d(x,z)\geq d(x,y)+d(y,z)
\end{equation}
\end{enumerate}

Hence we need a distance definition which satisfies the four rules and characterizes the seaport distance effectively. The diffusion distance \citep{Coifman2005Geometric} proposed by Segarra et al \cite{segarra2015diffusion} measures the distance between nodes on a network. It is a perfect choice based on all these considerations. Specifically, for a weighted network $G(E,V,W)$ with Laplacian matrix $L$ and node space $V$ having two signals $\textbf{r, s}\in\mathbb{R}^N$. In the diffusion process, the distribution of signal on this network can be express as a vector. The diffusion distance \cite{segarra2015diffusion} is defined as integral of the norm of two signals vector over time:
\begin{equation}
d_{diff}^L(\textbf{r, s})=\Arrowvert \int_{0}^{+\infty} e^{-t}e^{-\alpha Lt}(\textbf{r}-\textbf{s}) dt \Arrowvert
\end{equation}
where $L=D-A$, with $D$ a diagonal matrix with its diagonal elements being the vertex degrees, $A$ is adjacency matrix, $I$ is identify matrix, $\alpha$ is diffusion constant, and $\Arrowvert\bullet \Arrowvert$ is defined norm. This equation equivalent to
\begin{equation}
d_{diff}^L(\textbf{r, s})=\Arrowvert (I+\alpha L)^{-1}(\textbf{r}-\textbf{s}) \Arrowvert
\end{equation}

Next, we need to prove that the diffusion distance fits into the metric space. The first three rules are straightforward. The fourth rule was proved follows: For $\textbf{v},\textbf{w}\in\mathbb{R}^N$,
\begin{equation}
\begin{aligned}
{\Arrowvert \textbf{v}+\textbf{w} \Arrowvert}_{diff}^L&=\Arrowvert (I+\alpha L)^{-1}(\textbf{v}+\textbf{w}) \Arrowvert\\
&\leq \Arrowvert (I+\alpha L)^{-1}\textbf{v} \Arrowvert +\Arrowvert (I+\alpha L)^{-1}\textbf{w} \Arrowvert\\
&={\Arrowvert \textbf{v} \Arrowvert}_{diff}^L+{\Arrowvert \textbf{w} \Arrowvert}_{diff}^L
\end{aligned}
\end{equation}

\begin{figure}[htbp]
\centering
\includegraphics[width=0.45\textwidth]{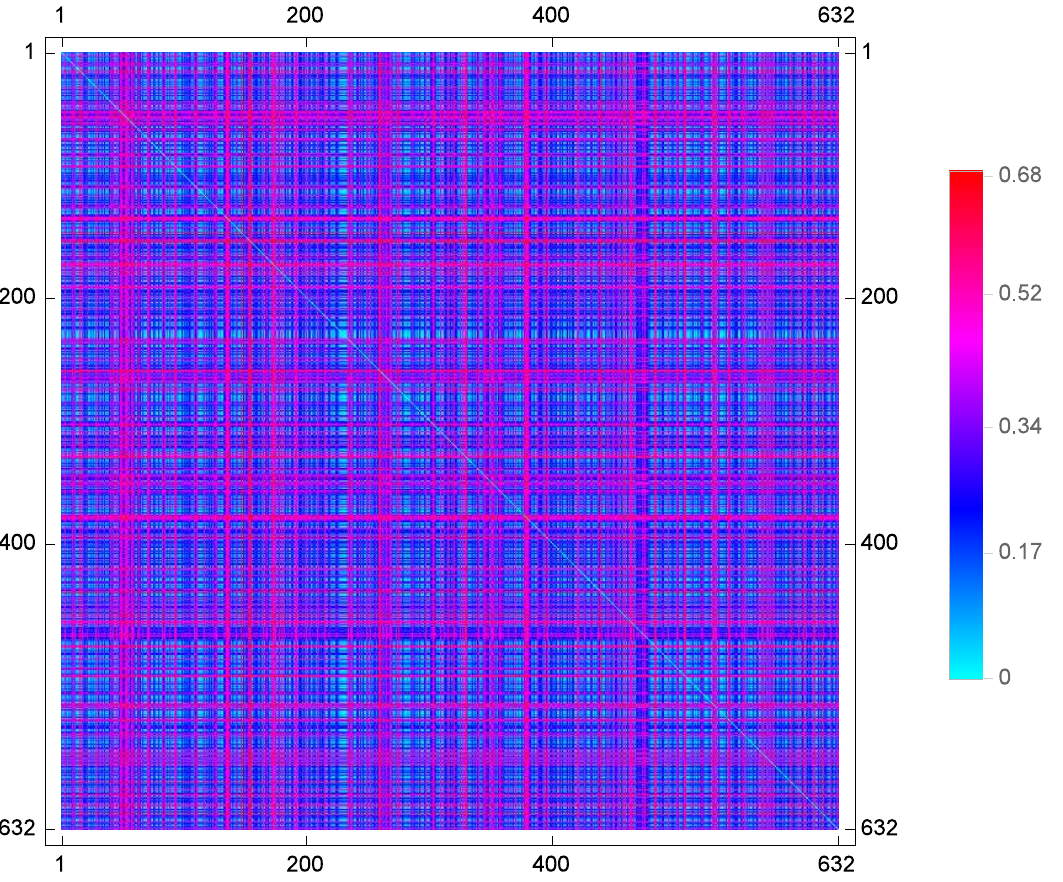}
\caption{(Color online) Heat graph of diffusion distances of 632 seaports of WMTN. The colors of blockage represent the diffusion distances between corresponding seaports, the hotter the color the further the distance.}
\label{fig:distance}
\end{figure}

\begin{figure*}[htbp]
\centering
\includegraphics[width=0.95\textwidth]{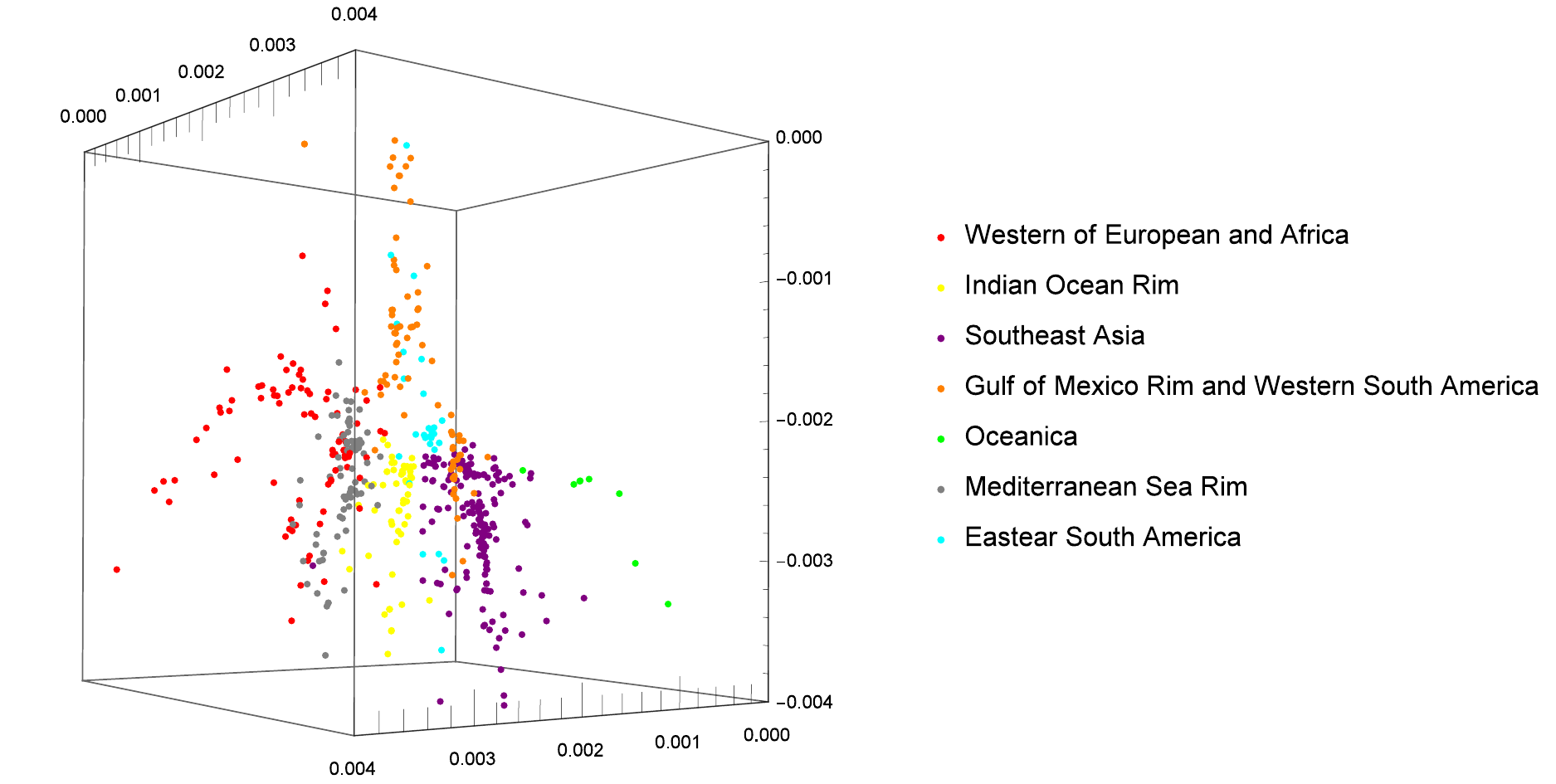}
\caption{(Color online) The 3-dimensional scaling graph of WMTN. Each node represents one seaport and the seaports with the same color belong to one community. Tinctorial representation is the same as in Fig.~\ref{fig:community}.}
\label{fig:mds}
\end{figure*}

The diffusion distances between each pair of nodes are presented in the distance matrix, and also shown in Fig.~\ref{fig:distance}. Then the MDS graph of WMTN can be obtained from the distance matrix. As shown in Fig.~\ref{fig:mds}, all the seaports within one community tend to gather into one cluster in the MDS graph. This phenomenon shows that the community structure detected by the LPAF is convincing. Also, the MDS is very intuitive and can be complementary to the modularity method in testing the result of community structure detection.
\section{Multilayer networks of WMTN}

\begin{figure}[htbp]
\centering
\includegraphics[width=9.4cm]{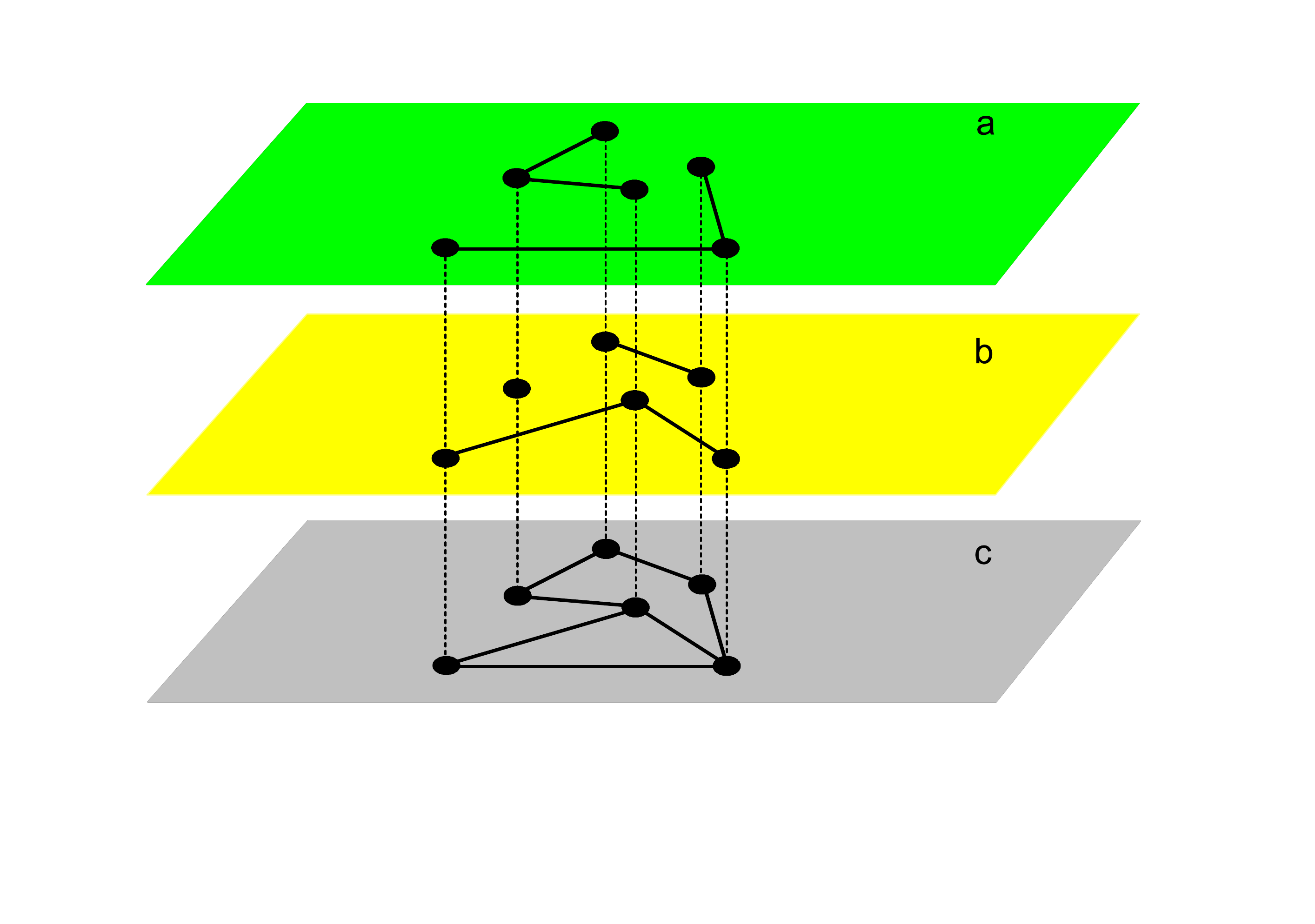}
\caption{(Color online) A simple means to aggregate different layers of networks. The three layers have the same nodes, and the edges in layer c contains all edges of layers a and b \cite{Domenico2015MuxViz}.}
\label{fig:aggregated}
\end{figure}

The development of network theory brings in more aspects of the network into consideration. For instance, temporal networks \cite{holme2012temporal} consider the time attribute for edges, and the multilayer networks \cite{Domenico2016The,kivela2014multilayer} consider the properties of different layers. Multilayer networks research can be classified into two categories \cite{Bianconi2015Interdisciplinary}. One is multiplex networks in which different layered networks have the same nodes and different types of connections. The another one is called network of networks \cite{Gao2012Networks,ferreira2014networks}in which different layered networks are constituted by different nodes and there are some interactions among this nodes. The concept of multilayer network is still under development but receives a great deal of attention. One simple method for studying multilayer network is, for the moment, converting multilayer networks to common networks. As shown in Fig.~\ref{fig:aggregated}, for unweighted networks $G(E,V)$ the graph layers a and b have the same nodes but different edges. It is called aggregate process, and the graph layer c contains all edges information of graph layers a and b.
\begin{equation}
G_c(E_c,V)=G_a(E_a,V)\cup G_b(E_b,V)
\end{equation}
The above strategy is too simple and may cause some problems. The most significant problem is the loss of relevant information. All the inter-layer information is removed during the aggregation. We can see in Fig.~\ref{fig:aggregated} that layer c can be aggregated by layers a and b, but layers a and b cannot be derived from layer c. However, the aggregation of the WMTN is less affected by the loss of inter-layer information, because different layers represent different marine companies, whose connection information is very similar.

\begin{figure*}[htbp]
\centering
\includegraphics[width=0.95\textwidth]{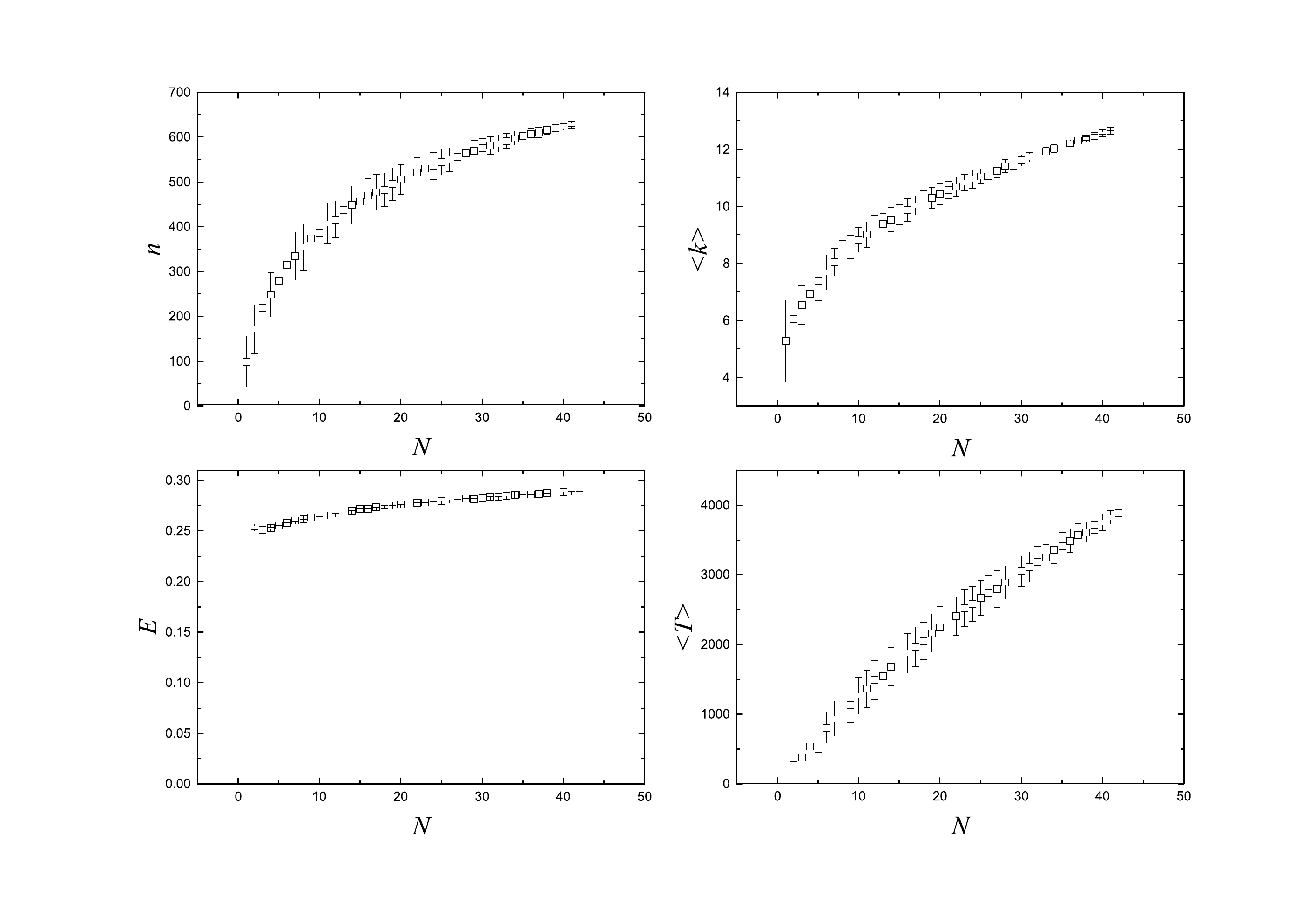}
\caption{Nodes number $n$, average degree $\left<K\right>$, networks efficiency $E$ and ATT $\left<T\right>$ versus the number $N$ of marine companies aggregated. Each single point is an average over 410 different realizations. Error bars are also provided.}
\label{fig:companies aggregated}
\end{figure*}

We studied the redundancy of different marine companies by multilayer network. If several companies have high redundancy implies that they may share very large operate area within a common sea area. Each marine company's data can be mapped onto one layered network, which is chosen randomly to be aggregated. Then we study the variation of nodes number, average degree, network efficiency and average trapping time(ATT) \cite{Montroll1969Random,lin2013random} versus the number of marine companies aggregated. The ATT can regarded as a measure the average of $F_{ij}$ which is the average first arriving time for a random walk from node $i$ to $j$. This quantity can describe transport efficiency and strongly depends on networks structure. As shown in Fig.~\ref{fig:companies aggregated}, nodes number and average degree does not reach the limits. This can be explained by the fact that the overlap ratio between different marine companies is low, and the current marine transport market can accommodate all existing marine companies. In addition, the efficiency of the WMTN is growing very slowly, and the ATT of the WMTN is almost linearly correlated with the number of marine companies aggregated. This result means that the WMTN has no structure breakdown in the aggregation process. So the structures of different marine companies are similar.

\begin{figure*}[htbp]
\centering
\includegraphics[width=1\textwidth]{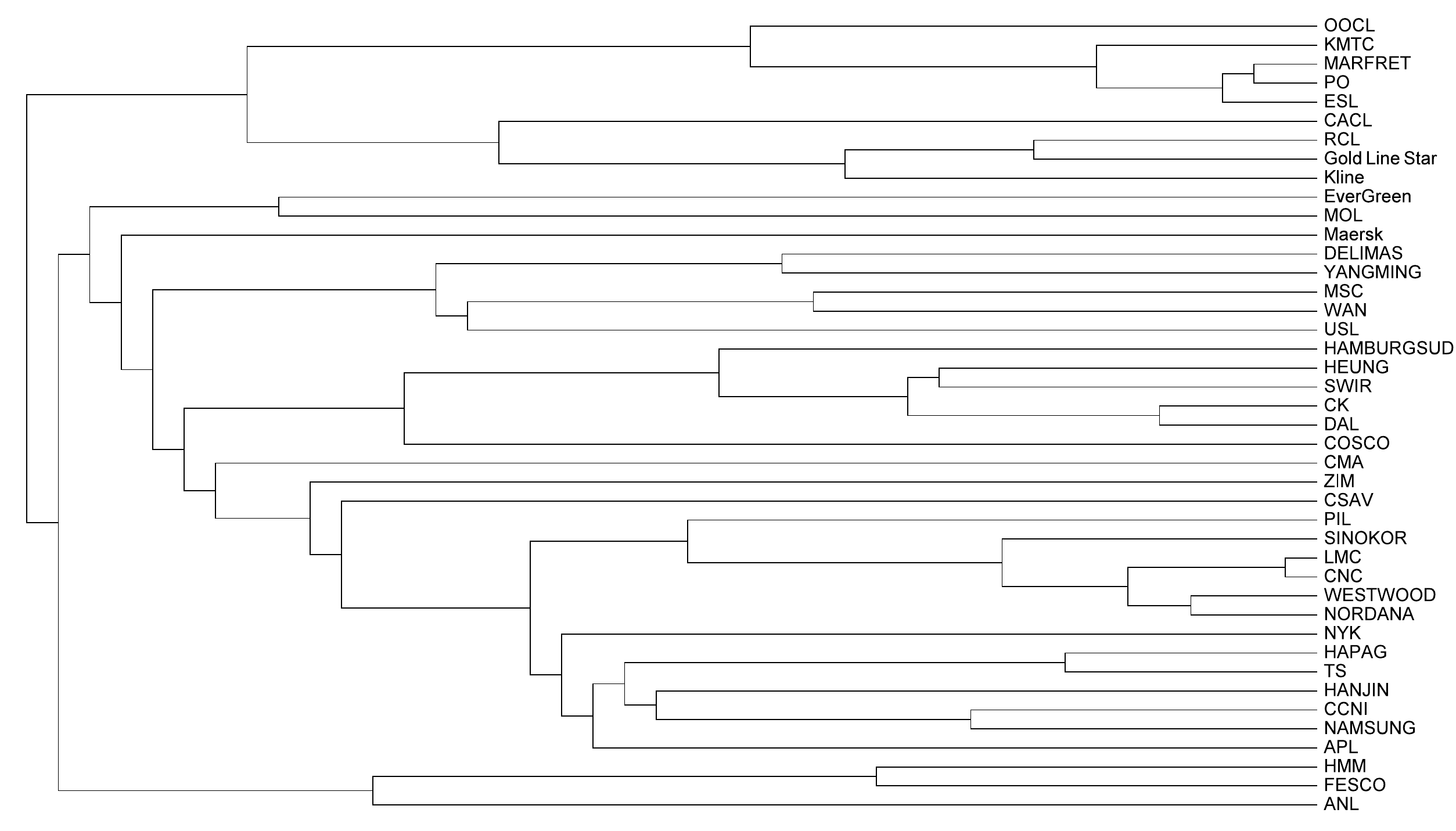}
\caption{The dendrogram of marine companies aggregations sequence. The sequence of aggregations is from right to left, and two child nodes aggregated to the one parent node.}
\label{fig:entropy dendrogram}
\end{figure*}

The structure reducibility of multilayer networks was studied in \cite{De2015Structural}. Manlio De Domenico1 et.al. have show that some layer information in multilayer might be redundant. And the main process of this methodology was show in following. Now we study the structure reducibility of WMTN. We aggregate the marine companies orderly, instead of random, by two shipping marine companies with minimum Jensen-Shannon distances showed \citep{Lin1991Divergence} in Fig.~\ref{fig:entropy dendrogram}. The Jensen-Shannon distance is a quantity to measure the similarity of different series, and is useful in complex system field \cite{Masucci2011Extracting,Gerlach2015Similarity}. For two networks with Laplacian matrix eigenvalue sequences $P(x)$ and $Q(x)$, respectively, which are serialized by Gaussian kernel $g(x, \gamma)=\frac{1}{\sqrt{2\pi \sigma^2}}exp(-\frac{(x-\gamma)^2}{2\sigma^2})$, the Jensen-Shannon distance can be expressed as:
\begin{widetext}
\begin{equation}
JS(P,Q)=\frac{1}{2}\left(\int P(x)\log (\frac{P(x)}{R(x)})\,dx+ \int Q(x)\log (\frac{Q(x)}{R(x)})\,dx\right),
\end{equation}
\end{widetext}
where $R=\frac{1}{2}(P+Q)$.

Firstly, we calculate Jensen-Shannon distances between every pair of marine companies set. Then, the networks of two marine companies with the minimum Jensen-Shannon distance will be aggregated to a new network, which is subject to a new round of aggregation. Eventually 42 networks sets can be obtained. Fig.~\ref{fig:entropy dendrogram} is the dendrogram \cite{EverittThe} of marine companies networks aggregation, which can illustrate the aggregation of the companies in a sequential manner. Such as, the aggregated order is from right to left, and two child nodes aggregated to the parent node in every step, in the first aggregated step, LMC and CNC are aggregated together.

\begin{figure}[htbp]
\centering
\includegraphics[width=0.5\textwidth]{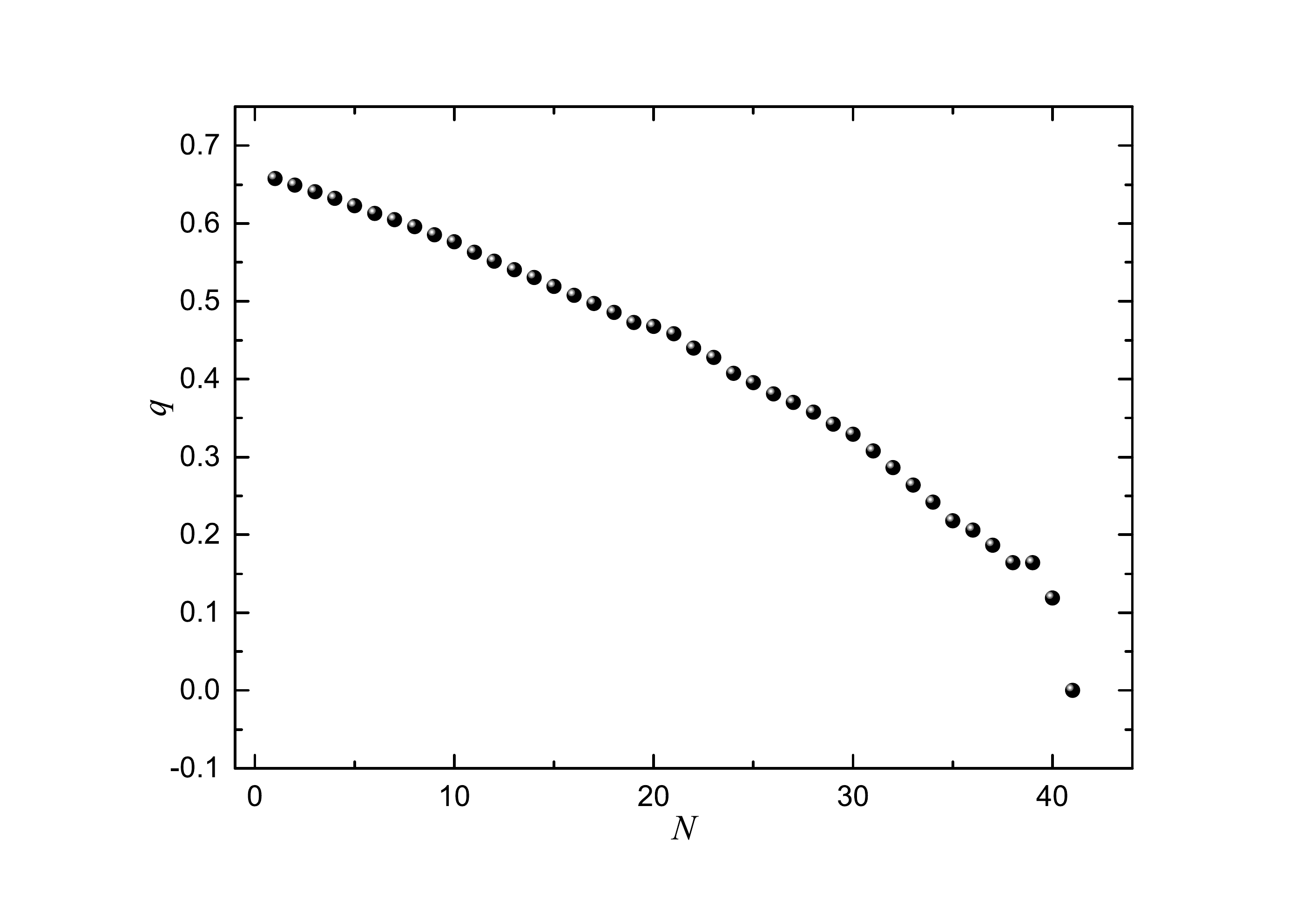}
\caption{The relative entropy $q$ versus the number $N$ of marine companies. It decreases smoothly with the increase of $N$}
\label{fig:relative entropy}
\end{figure}

For a network with $N$ nodes and normalized Laplacian matrix $L_G$, the von Neumann entropy of this network is:
\begin{equation}
h=-\sum_{i=1}^N\lambda_i\log_2(\lambda_i),
\end{equation}
where $\{\lambda_i\}$ is the eigenvalue of $L_G$.
In the aggregation process, every networks sets can be calculated by relative entropy $q(c)=1-\frac{\overline{h(c)}}{h_A}$, where $\overline{h(c)}=\frac{\sum_{\alpha=1}h_{C^[\alpha]}}{X}$ and $C$ is the aggregation step, $X$ is the number of networks in $C$-th step, and $h_A$ is the von Neumann entropy of the network with all the layers aggregated. For a multilayer network $\mathcal{A}$, Its reducibility is defined as\cite{De2015Structural}
\begin{equation}
\mathcal{X}=\frac{M-M_{opt}}{M-1}
\end{equation}
where $M$ the amount of layer, $M_{opt}$ is the number of layer while $q(c)$ have max value in the aggregated process.
The result is shown in Fig.~\ref{fig:relative entropy}, suggesting the relative entropy decreases smoothly with the number of marine companies aggregated. The relative entropy even keeps decreasing at the stage of all marine companies aggregating to several marine companies. So the value of reducibility of WMTN is 1. Such a phenomenon indicates that all marine companies are relatively independent and the overlap between them is low \cite{De2015Structural}. This also shows that the redundancy of WMTN is low, with each marine company having its local dominating area.

\section{Conclusion}

In this paper, we calculated the topological properties of WMTN and compare it with Poland city bus networks. The WMTN have higher networks efficiency and higher degree. we proposed the line saturability to measure the line transport ability, and the result is 0.877. Otherwise, we found the line saturability decreases exponentially with the length of lines. The WMTN can be divided into 7 communities by the LPAF, which is in agreement with the MDS graph. The seaports of one community are distributed in the same geographical area on world map. It mean the WMTN was limited by geography, the transport of local is more frequent than global. We also studied the relation between marine companies by means of multilayer networks. The different marine companies have low overlap and similar network structures. All of the results above show that the marine transport has high efficiency and low redundancy rate.

\section*{Acknowledgements}
We gratefully acknowledge the fruitful discussions with Shengfeng Deng, Longfeng Zhao, Jihui Han, Liang Zheng and Ke Wang. This work was supported in part by National Natural Science Foundation of China (Grant No. 11505071, 11647048), the Programme of Introducing Talents of Discipline to Universities under Grant No. B08033 the Fundamental Research Funds for the Central Universities (Grant No. KJ02072016-0170, CCNU), and the Research Project of Hubei Provincial Department of Education No. Q20161703.

\section*{References}

\bibliography{apssamp}

\begin{thebibliography}{42}%
\makeatletter
\providecommand \@ifxundefined [1]{%
 \@ifx{#1\undefined}
}%
\providecommand \@ifnum [1]{%
 \ifnum #1\expandafter \@firstoftwo
 \else \expandafter \@secondoftwo
 \fi
}%
\providecommand \@ifx [1]{%
 \ifx #1\expandafter \@firstoftwo
 \else \expandafter \@secondoftwo
 \fi
}%
\providecommand \natexlab [1]{#1}%
\providecommand \enquote  [1]{``#1''}%
\providecommand \bibnamefont  [1]{#1}%
\providecommand \bibfnamefont [1]{#1}%
\providecommand \citenamefont [1]{#1}%
\providecommand \href@noop [0]{\@secondoftwo}%
\providecommand \href [0]{\begingroup \@sanitize@url \@href}%
\providecommand \@href[1]{\@@startlink{#1}\@@href}%
\providecommand \@@href[1]{\endgroup#1\@@endlink}%
\providecommand \@sanitize@url [0]{\catcode `\\12\catcode `\$12\catcode
  `\&12\catcode `\#12\catcode `\^12\catcode `\_12\catcode `\%12\relax}%
\providecommand \@@startlink[1]{}%
\providecommand \@@endlink[0]{}%
\providecommand \url  [0]{\begingroup\@sanitize@url \@url }%
\providecommand \@url [1]{\endgroup\@href {#1}{\urlprefix }}%
\providecommand \urlprefix  [0]{URL }%
\providecommand \Eprint [0]{\href }%
\providecommand \doibase [0]{https://doi.org/}%
\providecommand \selectlanguage [0]{\@gobble}%
\providecommand \bibinfo  [0]{\@secondoftwo}%
\providecommand \bibfield  [0]{\@secondoftwo}%
\providecommand \translation [1]{[#1]}%
\providecommand \BibitemOpen [0]{}%
\providecommand \bibitemStop [0]{}%
\providecommand \bibitemNoStop [0]{.\EOS\space}%
\providecommand \EOS [0]{\spacefactor3000\relax}%
\providecommand \BibitemShut  [1]{\csname bibitem#1\endcsname}%
\let\auto@bib@innerbib\@empty
\bibitem [{\citenamefont {unctad}()}]{unctad}%
  \BibitemOpen
  \bibfield  {author} {\bibinfo {author} {\bibnamefont {unctad}},\ }\bibfield
  {title} {\bibinfo {title} {{marine transportation}},\ }\href@noop {} {\
  }\bibinfo {note}
  {\url{http://unctad.org/en/pages/PublicationWebflyer.aspx?publicationid=1374}}\BibitemShut
  {NoStop}%
\bibitem [{\citenamefont {Haezendonck}\ \emph {et~al.}(2000)\citenamefont
  {Haezendonck}, \citenamefont {Pison}, \citenamefont {Rousseeuw},
  \citenamefont {Struyf}, \citenamefont {Verbeke} \emph
  {et~al.}}]{haezendonck2000competitive}%
  \BibitemOpen
  \bibfield  {author} {\bibinfo {author} {\bibfnamefont {E.}~\bibnamefont
  {Haezendonck}}, \bibinfo {author} {\bibfnamefont {G.}~\bibnamefont {Pison}},
  \bibinfo {author} {\bibfnamefont {P.}~\bibnamefont {Rousseeuw}}, \bibinfo
  {author} {\bibfnamefont {A.}~\bibnamefont {Struyf}}, \bibinfo {author}
  {\bibfnamefont {A.}~\bibnamefont {Verbeke}}, \emph {et~al.},\ }\bibfield
  {title} {\bibinfo {title} {The competitive advantage of seaports},\
  }\href@noop {} {\bibfield  {journal} {\bibinfo  {journal} {International
  Journal of Maritime Economics}\ }\textbf {\bibinfo {volume} {2}},\ \bibinfo
  {pages} {69} (\bibinfo {year} {2000})}\BibitemShut {NoStop}%
\bibitem [{\citenamefont {Li}\ and\ \citenamefont
  {Cai}(2004)}]{Li2004Statistical}%
  \BibitemOpen
  \bibfield  {author} {\bibinfo {author} {\bibfnamefont {W.}~\bibnamefont
  {Li}}\ and\ \bibinfo {author} {\bibfnamefont {X.}~\bibnamefont {Cai}},\
  }\bibfield  {title} {\bibinfo {title} {Statistical analysis of airport
  network of china},\ }\href@noop {} {\bibfield  {journal} {\bibinfo  {journal}
  {Physical Review E}\ }\textbf {\bibinfo {volume} {69}},\ \bibinfo {pages}
  {046106} (\bibinfo {year} {2004})}\BibitemShut {NoStop}%
\bibitem [{\citenamefont {Hu}\ \emph {et~al.}(2006)\citenamefont {Hu},
  \citenamefont {Wang}, \citenamefont {Jiang}, \citenamefont {Wu},
  \citenamefont {Wang},\ and\ \citenamefont {Wu}}]{Hu2006Urban}%
  \BibitemOpen
  \bibfield  {author} {\bibinfo {author} {\bibfnamefont {M.~B.}\ \bibnamefont
  {Hu}}, \bibinfo {author} {\bibfnamefont {W.~X.}\ \bibnamefont {Wang}},
  \bibinfo {author} {\bibfnamefont {R.}~\bibnamefont {Jiang}}, \bibinfo
  {author} {\bibfnamefont {Q.~S.}\ \bibnamefont {Wu}}, \bibinfo {author}
  {\bibfnamefont {B.~H.}\ \bibnamefont {Wang}},\ and\ \bibinfo {author}
  {\bibfnamefont {Y.~H.}\ \bibnamefont {Wu}},\ }\bibfield  {title} {\bibinfo
  {title} {Urban traffic dynamics: A scale-free network perspective},\
  }\href@noop {} {\bibfield  {journal} {\bibinfo  {journal} {Physics}\ }
  (\bibinfo {year} {2006})}\BibitemShut {NoStop}%
\bibitem [{\citenamefont {Puffert}(2002)}]{Puffert2002Path}%
  \BibitemOpen
  \bibfield  {author} {\bibinfo {author} {\bibfnamefont {D.~J.}\ \bibnamefont
  {Puffert}},\ }\bibfield  {title} {\bibinfo {title} {Path dependence in
  spatial networks: The standardization of railway track gauge},\ }\href@noop
  {} {\bibfield  {journal} {\bibinfo  {journal} {Explorations in Economic
  History}\ }\textbf {\bibinfo {volume} {39}},\ \bibinfo {pages} {282}
  (\bibinfo {year} {2002})}\BibitemShut {NoStop}%
\bibitem [{\citenamefont {Mart{\'\i}-Henneberg}(2013)}]{marti2013european}%
  \BibitemOpen
  \bibfield  {author} {\bibinfo {author} {\bibfnamefont {J.}~\bibnamefont
  {Mart{\'\i}-Henneberg}},\ }\bibfield  {title} {\bibinfo {title} {European
  integration and national models for railway networks (1840--2010)},\
  }\href@noop {} {\bibfield  {journal} {\bibinfo  {journal} {Journal of
  Transport Geography}\ }\textbf {\bibinfo {volume} {26}},\ \bibinfo {pages}
  {126} (\bibinfo {year} {2013})}\BibitemShut {NoStop}%
\bibitem [{\citenamefont {Wei-Bing}\ \emph {et~al.}(2009)\citenamefont
  {Wei-Bing}, \citenamefont {Long}, \citenamefont {Wei},\ and\ \citenamefont
  {Xu}}]{wei2009worldwide}%
  \BibitemOpen
  \bibfield  {author} {\bibinfo {author} {\bibfnamefont {D.}~\bibnamefont
  {Wei-Bing}}, \bibinfo {author} {\bibfnamefont {G.}~\bibnamefont {Long}},
  \bibinfo {author} {\bibfnamefont {L.}~\bibnamefont {Wei}},\ and\ \bibinfo
  {author} {\bibfnamefont {C.}~\bibnamefont {Xu}},\ }\bibfield  {title}
  {\bibinfo {title} {Worldwide marine transportation network: Efficiency and
  container throughput},\ }\href@noop {} {\bibfield  {journal} {\bibinfo
  {journal} {Chinese Physics Letters}\ }\textbf {\bibinfo {volume} {26}},\
  \bibinfo {pages} {118901} (\bibinfo {year} {2009})}\BibitemShut {NoStop}%
\bibitem [{\citenamefont {Kaluza}\ \emph {et~al.}(2010)\citenamefont {Kaluza},
  \citenamefont {K{\"o}lzsch}, \citenamefont {Gastner},\ and\ \citenamefont
  {Blasius}}]{kaluza2010complex}%
  \BibitemOpen
  \bibfield  {author} {\bibinfo {author} {\bibfnamefont {P.}~\bibnamefont
  {Kaluza}}, \bibinfo {author} {\bibfnamefont {A.}~\bibnamefont {K{\"o}lzsch}},
  \bibinfo {author} {\bibfnamefont {M.~T.}\ \bibnamefont {Gastner}},\ and\
  \bibinfo {author} {\bibfnamefont {B.}~\bibnamefont {Blasius}},\ }\bibfield
  {title} {\bibinfo {title} {The complex network of global cargo ship
  movements},\ }\href@noop {} {\bibfield  {journal} {\bibinfo  {journal}
  {Journal of the Royal Society Interface}\ }\textbf {\bibinfo {volume} {7}},\
  \bibinfo {pages} {1093} (\bibinfo {year} {2010})}\BibitemShut {NoStop}%
\bibitem [{\citenamefont {Ducruet}\ and\ \citenamefont
  {Notteboom}(2012)}]{ducruet2012worldwide}%
  \BibitemOpen
  \bibfield  {author} {\bibinfo {author} {\bibfnamefont {C.}~\bibnamefont
  {Ducruet}}\ and\ \bibinfo {author} {\bibfnamefont {T.}~\bibnamefont
  {Notteboom}},\ }\bibfield  {title} {\bibinfo {title} {The worldwide maritime
  network of container shipping: spatial structure and regional dynamics},\
  }\href@noop {} {\bibfield  {journal} {\bibinfo  {journal} {Global Networks}\
  }\textbf {\bibinfo {volume} {12}},\ \bibinfo {pages} {395} (\bibinfo {year}
  {2012})}\BibitemShut {NoStop}%
\bibitem [{\citenamefont {Han}\ \emph {et~al.}(2016)\citenamefont {Han},
  \citenamefont {Li}, \citenamefont {Su}, \citenamefont {Zhao},\ and\
  \citenamefont {Deng}}]{Han2016Community}%
  \BibitemOpen
  \bibfield  {author} {\bibinfo {author} {\bibfnamefont {J.}~\bibnamefont
  {Han}}, \bibinfo {author} {\bibfnamefont {W.}~\bibnamefont {Li}}, \bibinfo
  {author} {\bibfnamefont {Z.}~\bibnamefont {Su}}, \bibinfo {author}
  {\bibfnamefont {L.}~\bibnamefont {Zhao}},\ and\ \bibinfo {author}
  {\bibfnamefont {W.}~\bibnamefont {Deng}},\ }\bibfield  {title} {\bibinfo
  {title} {Community detection by label propagation with compression of flow},\
  }\href@noop {} {\bibfield  {journal} {\bibinfo  {journal} {European Physical
  Journal B}\ }\textbf {\bibinfo {volume} {89}},\ \bibinfo {pages} {272}
  (\bibinfo {year} {2016})}\BibitemShut {NoStop}%
\bibitem [{\citenamefont {Borg}\ and\ \citenamefont
  {Groenen}(2005)}]{borg2005modern}%
  \BibitemOpen
  \bibfield  {author} {\bibinfo {author} {\bibfnamefont {I.}~\bibnamefont
  {Borg}}\ and\ \bibinfo {author} {\bibfnamefont {P.~J.}\ \bibnamefont
  {Groenen}},\ }\href@noop {} {\emph {\bibinfo {title} {Modern multidimensional
  scaling: Theory and applications}}}\ (\bibinfo  {publisher} {Springer Science
  \& Business Media},\ \bibinfo {year} {2005})\BibitemShut {NoStop}%
\bibitem [{\citenamefont {Segarra}\ \emph {et~al.}(2015)\citenamefont
  {Segarra}, \citenamefont {Huang},\ and\ \citenamefont
  {Ribeiro}}]{segarra2015diffusion}%
  \BibitemOpen
  \bibfield  {author} {\bibinfo {author} {\bibfnamefont {S.}~\bibnamefont
  {Segarra}}, \bibinfo {author} {\bibfnamefont {W.}~\bibnamefont {Huang}},\
  and\ \bibinfo {author} {\bibfnamefont {A.}~\bibnamefont {Ribeiro}},\
  }\bibfield  {title} {\bibinfo {title} {Diffusion and superposition distances
  for signals supported on networks},\ }\href@noop {} {\bibfield  {journal}
  {\bibinfo  {journal} {Signal and Information Processing over Networks, IEEE
  Transactions on}\ }\textbf {\bibinfo {volume} {1}},\ \bibinfo {pages} {20}
  (\bibinfo {year} {2015})}\BibitemShut {NoStop}%
\bibitem [{\citenamefont {Lin}(1991)}]{Lin1991Divergence}%
  \BibitemOpen
  \bibfield  {author} {\bibinfo {author} {\bibfnamefont {J.}~\bibnamefont
  {Lin}},\ }\href@noop {} {\emph {\bibinfo {title} {Divergence measures based
  on the Shannon entropy}}}\ (\bibinfo  {publisher} {IEEE Press},\ \bibinfo
  {year} {1991})\ pp.\ \bibinfo {pages} {145--151}\BibitemShut {NoStop}%
\bibitem [{\citenamefont {Domenico}\ and\ \citenamefont
  {Biamonte}(2016)}]{Domenico2016Spectral}%
  \BibitemOpen
  \bibfield  {author} {\bibinfo {author} {\bibfnamefont {M.~D.}\ \bibnamefont
  {Domenico}}\ and\ \bibinfo {author} {\bibfnamefont {J.}~\bibnamefont
  {Biamonte}},\ }\bibfield  {title} {\bibinfo {title} {Spectral entropies as
  information-theoretic tools for complex network comparison},\ }\href@noop {}
  {\bibfield  {journal} {\bibinfo  {journal} {Physical Review X}\ }\textbf
  {\bibinfo {volume} {6}} (\bibinfo {year} {2016})}\BibitemShut {NoStop}%
\bibitem [{\citenamefont {De}\ \emph {et~al.}(2015)\citenamefont {De},
  \citenamefont {Nicosia}, \citenamefont {Arenas},\ and\ \citenamefont
  {Latora}}]{De2015Structural}%
  \BibitemOpen
  \bibfield  {author} {\bibinfo {author} {\bibfnamefont {D.~M.}\ \bibnamefont
  {De}}, \bibinfo {author} {\bibfnamefont {V.}~\bibnamefont {Nicosia}},
  \bibinfo {author} {\bibfnamefont {A.}~\bibnamefont {Arenas}},\ and\ \bibinfo
  {author} {\bibfnamefont {V.}~\bibnamefont {Latora}},\ }\bibfield  {title}
  {\bibinfo {title} {Structural reducibility of multilayer networks.},\
  }\href@noop {} {\bibfield  {journal} {\bibinfo  {journal} {Nature
  Communications}\ }\textbf {\bibinfo {volume} {6}} (\bibinfo {year}
  {2015})}\BibitemShut {NoStop}%
\bibitem [{\citenamefont {Feynman}\ and\ \citenamefont {{Vernon
  Jr.}}(1963)}]{Feynman1963118}%
  \BibitemOpen
  \bibfield  {author} {\bibinfo {author} {\bibfnamefont {R.}~\bibnamefont
  {Feynman}}\ and\ \bibinfo {author} {\bibfnamefont {F.}~\bibnamefont {{Vernon
  Jr.}}},\ }\bibfield  {title} {\bibinfo {title} {The theory of a general
  quantum system interacting with a linear dissipative system},\ }\href
  {https://doi.org/10.1016/0003-4916(63)90068-X} {\bibfield  {journal}
  {\bibinfo  {journal} {Annals of Physics}\ }\textbf {\bibinfo {volume} {24}},\
  \bibinfo {pages} {118} (\bibinfo {year} {1963})}\BibitemShut {NoStop}%
\bibitem [{\citenamefont {COSCO}()}]{COSCO2014}%
  \BibitemOpen
  \bibfield  {author} {\bibinfo {author} {\bibnamefont {COSCO}},\ }\bibfield
  {title} {\bibinfo {title} {{COSCO shipping lines}},\ }\href@noop {} {\
  }\bibinfo {note}
  {\url{http://lines.coscoshipping.com/ourservice/toService.do}}\BibitemShut
  {NoStop}%
\bibitem [{\citenamefont {ANL}()}]{ANL2014}%
  \BibitemOpen
  \bibfield  {author} {\bibinfo {author} {\bibnamefont {ANL}},\ }\bibfield
  {title} {\bibinfo {title} {{ANL shipping lines}},\ }\href@noop {} {\
  }\bibinfo {note}
  {\url{http://www.anl.com.au/products-services/line-services/solution.}}\BibitemShut
  {Stop}%
\bibitem [{\citenamefont {APL}()}]{APL2014}%
  \BibitemOpen
  \bibfield  {author} {\bibinfo {author} {\bibnamefont {APL}},\ }\bibfield
  {title} {\bibinfo {title} {{APL shipping lines}},\ }\href@noop {} {\
  }\bibinfo {note}
  {\url{http://www.apl.com/wps/portal/apl/apl-home/services/OCEANALLIANCE}}\BibitemShut
  {NoStop}%
\bibitem [{\citenamefont {HNM}()}]{HNM2014}%
  \BibitemOpen
  \bibfield  {author} {\bibinfo {author} {\bibnamefont {HNM}},\ }\bibfield
  {title} {\bibinfo {title} {{HNM shipping lines}},\ }\href@noop {} {\
  }\bibinfo {note}
  {\url{http://www.hmm.co.kr/cms/company/engn/container/service/index.jsp}}\BibitemShut
  {NoStop}%
\bibitem [{\citenamefont {Sen}\ \emph {et~al.}(2003)\citenamefont {Sen},
  \citenamefont {Dasgupta}, \citenamefont {Chatterjee}, \citenamefont
  {Sreeram}, \citenamefont {Mukherjee},\ and\ \citenamefont
  {Manna}}]{sen2003small}%
  \BibitemOpen
  \bibfield  {author} {\bibinfo {author} {\bibfnamefont {P.}~\bibnamefont
  {Sen}}, \bibinfo {author} {\bibfnamefont {S.}~\bibnamefont {Dasgupta}},
  \bibinfo {author} {\bibfnamefont {A.}~\bibnamefont {Chatterjee}}, \bibinfo
  {author} {\bibfnamefont {P.}~\bibnamefont {Sreeram}}, \bibinfo {author}
  {\bibfnamefont {G.}~\bibnamefont {Mukherjee}},\ and\ \bibinfo {author}
  {\bibfnamefont {S.}~\bibnamefont {Manna}},\ }\bibfield  {title} {\bibinfo
  {title} {Small-world properties of the indian railway network},\ }\href@noop
  {} {\bibfield  {journal} {\bibinfo  {journal} {Physical Review E}\ }\textbf
  {\bibinfo {volume} {67}},\ \bibinfo {pages} {036106} (\bibinfo {year}
  {2003})}\BibitemShut {NoStop}%
\bibitem [{\citenamefont {Latora}\ and\ \citenamefont
  {Marchiori}(2001)}]{Latora2001Efficient}%
  \BibitemOpen
  \bibfield  {author} {\bibinfo {author} {\bibfnamefont {V.}~\bibnamefont
  {Latora}}\ and\ \bibinfo {author} {\bibfnamefont {M.}~\bibnamefont
  {Marchiori}},\ }\bibfield  {title} {\bibinfo {title} {Efficient behavior of
  small-world networks},\ }\href@noop {} {\bibfield  {journal} {\bibinfo
  {journal} {Physical Review Letters}\ }\textbf {\bibinfo {volume} {87}},\
  \bibinfo {pages} {198701} (\bibinfo {year} {2001})}\BibitemShut {NoStop}%
\bibitem [{\citenamefont {Achard}\ and\ \citenamefont
  {Bullmore}(2007)}]{achard2007efficiency}%
  \BibitemOpen
  \bibfield  {author} {\bibinfo {author} {\bibfnamefont {S.}~\bibnamefont
  {Achard}}\ and\ \bibinfo {author} {\bibfnamefont {E.}~\bibnamefont
  {Bullmore}},\ }\bibfield  {title} {\bibinfo {title} {Efficiency and cost of
  economical brain functional networks},\ }\href@noop {} {\bibfield  {journal}
  {\bibinfo  {journal} {PLoS Comput Biol}\ }\textbf {\bibinfo {volume} {3}},\
  \bibinfo {pages} {e17} (\bibinfo {year} {2007})}\BibitemShut {NoStop}%
\bibitem [{\citenamefont {Sienkiewicz}\ and\ \citenamefont
  {Ho{\l}yst}(2005)}]{sienkiewicz2005statistical}%
  \BibitemOpen
  \bibfield  {author} {\bibinfo {author} {\bibfnamefont {J.}~\bibnamefont
  {Sienkiewicz}}\ and\ \bibinfo {author} {\bibfnamefont {J.~A.}\ \bibnamefont
  {Ho{\l}yst}},\ }\bibfield  {title} {\bibinfo {title} {Statistical analysis of
  22 public transport networks in poland},\ }\href@noop {} {\bibfield
  {journal} {\bibinfo  {journal} {Physical Review E}\ }\textbf {\bibinfo
  {volume} {72}},\ \bibinfo {pages} {046127} (\bibinfo {year}
  {2005})}\BibitemShut {NoStop}%
\bibitem [{\citenamefont {Watts}\ and\ \citenamefont
  {Strogatz}(1998)}]{Watts1998Collective}%
  \BibitemOpen
  \bibfield  {author} {\bibinfo {author} {\bibfnamefont {D.~J.}\ \bibnamefont
  {Watts}}\ and\ \bibinfo {author} {\bibfnamefont {S.~H.}\ \bibnamefont
  {Strogatz}},\ }\bibfield  {title} {\bibinfo {title} {Collective dynamics of
  'small-world' networks.},\ }\href@noop {} {\bibfield  {journal} {\bibinfo
  {journal} {Nature}\ }\textbf {\bibinfo {volume} {393}},\ \bibinfo {pages}
  {440} (\bibinfo {year} {1998})}\BibitemShut {NoStop}%
\bibitem [{\citenamefont {Berche}\ \emph {et~al.}(2009)\citenamefont {Berche},
  \citenamefont {Ferber}, \citenamefont {Holovatch},\ and\ \citenamefont
  {Holovatch}}]{Berche2009Resilience}%
  \BibitemOpen
  \bibfield  {author} {\bibinfo {author} {\bibfnamefont {B.}~\bibnamefont
  {Berche}}, \bibinfo {author} {\bibfnamefont {C.~V.}\ \bibnamefont {Ferber}},
  \bibinfo {author} {\bibfnamefont {T.}~\bibnamefont {Holovatch}},\ and\
  \bibinfo {author} {\bibfnamefont {Y.}~\bibnamefont {Holovatch}},\ }\bibfield
  {title} {\bibinfo {title} {Resilience of public transport networks against
  attacks},\ }\href@noop {} {\bibfield  {journal} {\bibinfo  {journal}
  {European Physical Journal B}\ }\textbf {\bibinfo {volume} {71}},\ \bibinfo
  {pages} {125} (\bibinfo {year} {2009})}\BibitemShut {NoStop}%
\bibitem [{\citenamefont {Girvan}\ and\ \citenamefont
  {Newman}(2002)}]{girvan2002community}%
  \BibitemOpen
  \bibfield  {author} {\bibinfo {author} {\bibfnamefont {M.}~\bibnamefont
  {Girvan}}\ and\ \bibinfo {author} {\bibfnamefont {M.~E.}\ \bibnamefont
  {Newman}},\ }\bibfield  {title} {\bibinfo {title} {Community structure in
  social and biological networks},\ }\href@noop {} {\bibfield  {journal}
  {\bibinfo  {journal} {Proceedings of the national academy of sciences}\
  }\textbf {\bibinfo {volume} {99}},\ \bibinfo {pages} {7821} (\bibinfo {year}
  {2002})}\BibitemShut {NoStop}%
\bibitem [{\citenamefont {Newman}(2004{\natexlab{a}})}]{newman2004detecting}%
  \BibitemOpen
  \bibfield  {author} {\bibinfo {author} {\bibfnamefont {M.~E.}\ \bibnamefont
  {Newman}},\ }\bibfield  {title} {\bibinfo {title} {Detecting community
  structure in networks},\ }\href@noop {} {\bibfield  {journal} {\bibinfo
  {journal} {The European Physical Journal B-Condensed Matter and Complex
  Systems}\ }\textbf {\bibinfo {volume} {38}},\ \bibinfo {pages} {321}
  (\bibinfo {year} {2004}{\natexlab{a}})}\BibitemShut {NoStop}%
\bibitem [{\citenamefont {Newman}(2004{\natexlab{b}})}]{newman2004fast}%
  \BibitemOpen
  \bibfield  {author} {\bibinfo {author} {\bibfnamefont {M.~E.}\ \bibnamefont
  {Newman}},\ }\bibfield  {title} {\bibinfo {title} {Fast algorithm for
  detecting community structure in networks},\ }\href@noop {} {\bibfield
  {journal} {\bibinfo  {journal} {Physical review E}\ }\textbf {\bibinfo
  {volume} {69}},\ \bibinfo {pages} {066133} (\bibinfo {year}
  {2004}{\natexlab{b}})}\BibitemShut {NoStop}%
\bibitem [{\citenamefont {Coifman}\ \emph {et~al.}(2005)\citenamefont
  {Coifman}, \citenamefont {Lafon}, \citenamefont {Lee}, \citenamefont
  {Maggioni}, \citenamefont {Nadler}, \citenamefont {Warner},\ and\
  \citenamefont {Zucker}}]{Coifman2005Geometric}%
  \BibitemOpen
  \bibfield  {author} {\bibinfo {author} {\bibfnamefont {R.~R.}\ \bibnamefont
  {Coifman}}, \bibinfo {author} {\bibfnamefont {S.}~\bibnamefont {Lafon}},
  \bibinfo {author} {\bibfnamefont {A.~B.}\ \bibnamefont {Lee}}, \bibinfo
  {author} {\bibfnamefont {M.}~\bibnamefont {Maggioni}}, \bibinfo {author}
  {\bibfnamefont {B.}~\bibnamefont {Nadler}}, \bibinfo {author} {\bibfnamefont
  {F.}~\bibnamefont {Warner}},\ and\ \bibinfo {author} {\bibfnamefont {S.~W.}\
  \bibnamefont {Zucker}},\ }\bibfield  {title} {\bibinfo {title} {Geometric
  diffusions as a tool for harmonic analysis and structure definition of data:
  Diffusion maps},\ }\href@noop {} {\bibfield  {journal} {\bibinfo  {journal}
  {Proceedings of the National Academy of Sciences of the United States of
  America}\ }\textbf {\bibinfo {volume} {102}},\ \bibinfo {pages} {7426}
  (\bibinfo {year} {2005})}\BibitemShut {NoStop}%
\bibitem [{\citenamefont {Holme}\ and\ \citenamefont
  {Saramäki}(2012)}]{holme2012temporal}%
  \BibitemOpen
  \bibfield  {author} {\bibinfo {author} {\bibfnamefont {P.}~\bibnamefont
  {Holme}}\ and\ \bibinfo {author} {\bibfnamefont {J.}~\bibnamefont
  {Saramäki}},\ }\bibfield  {title} {\bibinfo {title} {Temporal networks},\
  }\href@noop {} {\bibfield  {journal} {\bibinfo  {journal} {Physics Reports}\
  }\textbf {\bibinfo {volume} {519}},\ \bibinfo {pages} {97} (\bibinfo {year}
  {2012})}\BibitemShut {NoStop}%
\bibitem [{\citenamefont {Domenico}\ \emph {et~al.}(2016)\citenamefont
  {Domenico}, \citenamefont {Granell}, \citenamefont {Porter},\ and\
  \citenamefont {Arenas}}]{Domenico2016The}%
  \BibitemOpen
  \bibfield  {author} {\bibinfo {author} {\bibfnamefont {M.~D.}\ \bibnamefont
  {Domenico}}, \bibinfo {author} {\bibfnamefont {C.}~\bibnamefont {Granell}},
  \bibinfo {author} {\bibfnamefont {M.~A.}\ \bibnamefont {Porter}},\ and\
  \bibinfo {author} {\bibfnamefont {A.}~\bibnamefont {Arenas}},\ }\bibfield
  {title} {\bibinfo {title} {The physics of spreading processes in multilayer
  networks},\ }\href@noop {} {\bibfield  {journal} {\bibinfo  {journal} {Nature
  Physics}\ }\textbf {\bibinfo {volume} {12}} (\bibinfo {year}
  {2016})}\BibitemShut {NoStop}%
\bibitem [{\citenamefont {Kivel{\"a}}\ \emph {et~al.}(2014)\citenamefont
  {Kivel{\"a}}, \citenamefont {Arenas}, \citenamefont {Barthelemy},
  \citenamefont {Gleeson}, \citenamefont {Moreno},\ and\ \citenamefont
  {Porter}}]{kivela2014multilayer}%
  \BibitemOpen
  \bibfield  {author} {\bibinfo {author} {\bibfnamefont {M.}~\bibnamefont
  {Kivel{\"a}}}, \bibinfo {author} {\bibfnamefont {A.}~\bibnamefont {Arenas}},
  \bibinfo {author} {\bibfnamefont {M.}~\bibnamefont {Barthelemy}}, \bibinfo
  {author} {\bibfnamefont {J.~P.}\ \bibnamefont {Gleeson}}, \bibinfo {author}
  {\bibfnamefont {Y.}~\bibnamefont {Moreno}},\ and\ \bibinfo {author}
  {\bibfnamefont {M.~A.}\ \bibnamefont {Porter}},\ }\bibfield  {title}
  {\bibinfo {title} {Multilayer networks},\ }\href@noop {} {\bibfield
  {journal} {\bibinfo  {journal} {Journal of Complex Networks}\ }\textbf
  {\bibinfo {volume} {2}},\ \bibinfo {pages} {203} (\bibinfo {year}
  {2014})}\BibitemShut {NoStop}%
\bibitem [{\citenamefont {Bianconi}(2015)}]{Bianconi2015Interdisciplinary}%
  \BibitemOpen
  \bibfield  {author} {\bibinfo {author} {\bibfnamefont {G.}~\bibnamefont
  {Bianconi}},\ }\bibfield  {title} {\bibinfo {title} {Interdisciplinary and
  physics challenges of network theory},\ }\href@noop {} {\bibfield  {journal}
  {\bibinfo  {journal} {EPL (Europhysics Letters)}\ }\textbf {\bibinfo {volume}
  {111}},\ \bibinfo {pages} {56001} (\bibinfo {year} {2015})}\BibitemShut
  {NoStop}%
\bibitem [{\citenamefont {Gao}\ \emph {et~al.}(2012)\citenamefont {Gao},
  \citenamefont {Buldyrev}, \citenamefont {Stanley},\ and\ \citenamefont
  {Havlin}}]{Gao2012Networks}%
  \BibitemOpen
  \bibfield  {author} {\bibinfo {author} {\bibfnamefont {J.}~\bibnamefont
  {Gao}}, \bibinfo {author} {\bibfnamefont {S.~V.}\ \bibnamefont {Buldyrev}},
  \bibinfo {author} {\bibfnamefont {H.~E.}\ \bibnamefont {Stanley}},\ and\
  \bibinfo {author} {\bibfnamefont {S.}~\bibnamefont {Havlin}},\ }\bibfield
  {title} {\bibinfo {title} {Networks formed from interdependent networks},\
  }\href@noop {} {\bibfield  {journal} {\bibinfo  {journal} {Nature Physics}\
  }\textbf {\bibinfo {volume} {8}},\ \bibinfo {pages} {40} (\bibinfo {year}
  {2012})}\BibitemShut {NoStop}%
\bibitem [{\citenamefont {Ferreira}(2014)}]{ferreira2014networks}%
  \BibitemOpen
  \bibfield  {author} {\bibinfo {author} {\bibfnamefont {M.~A.~M.}\
  \bibnamefont {Ferreira}},\ }\bibfield  {title} {\bibinfo {title} {Networks of
  networks: The last frontier of complexity-a book review},\ }\href@noop {}
  {\bibfield  {journal} {\bibinfo  {journal} {International Journal of Latest
  Trends in Finance and Economic Sciences}\ }\textbf {\bibinfo {volume} {4}},\
  \bibinfo {pages} {708} (\bibinfo {year} {2014})}\BibitemShut {NoStop}%
\bibitem [{\citenamefont {Domenico}\ \emph {et~al.}(2015)\citenamefont
  {Domenico}, \citenamefont {Porter},\ and\ \citenamefont
  {Arenas}}]{Domenico2015MuxViz}%
  \BibitemOpen
  \bibfield  {author} {\bibinfo {author} {\bibfnamefont {M.~D.}\ \bibnamefont
  {Domenico}}, \bibinfo {author} {\bibfnamefont {M.~A.}\ \bibnamefont
  {Porter}},\ and\ \bibinfo {author} {\bibfnamefont {A.}~\bibnamefont
  {Arenas}},\ }\bibfield  {title} {\bibinfo {title} {Muxviz: a tool for
  multilayer analysis and visualization of networks},\ }\href@noop {}
  {\bibfield  {journal} {\bibinfo  {journal} {Jcomplexnetw}\ }\textbf {\bibinfo
  {volume} {3}} (\bibinfo {year} {2015})}\BibitemShut {NoStop}%
\bibitem [{\citenamefont {Montroll}(1969)}]{Montroll1969Random}%
  \BibitemOpen
  \bibfield  {author} {\bibinfo {author} {\bibfnamefont {E.~W.}\ \bibnamefont
  {Montroll}},\ }\bibfield  {title} {\bibinfo {title} {Random walks on
  lattices. iii. calculation of first‐passage times with application to
  exciton trapping on photosynthetic units},\ }\href@noop {} {\bibfield
  {journal} {\bibinfo  {journal} {Journal of Mathematical Physics}\ }\textbf
  {\bibinfo {volume} {10}},\ \bibinfo {pages} {753} (\bibinfo {year}
  {1969})}\BibitemShut {NoStop}%
\bibitem [{\citenamefont {Lin}\ \emph {et~al.}(2013)\citenamefont {Lin},
  \citenamefont {Zhang} \emph {et~al.}}]{lin2013random}%
  \BibitemOpen
  \bibfield  {author} {\bibinfo {author} {\bibfnamefont {Y.}~\bibnamefont
  {Lin}}, \bibinfo {author} {\bibfnamefont {Z.}~\bibnamefont {Zhang}}, \emph
  {et~al.},\ }\bibfield  {title} {\bibinfo {title} {Random walks in weighted
  networks with a perfect trap: An application of laplacian spectra},\
  }\href@noop {} {\bibfield  {journal} {\bibinfo  {journal} {Physical Review
  E}\ }\textbf {\bibinfo {volume} {87}},\ \bibinfo {pages} {062140} (\bibinfo
  {year} {2013})}\BibitemShut {NoStop}%
\bibitem [{\citenamefont {Masucci}\ \emph {et~al.}(2011)\citenamefont
  {Masucci}, \citenamefont {Kalampokis}, \citenamefont {Eguíluz},\ and\
  \citenamefont {Hernández-García}}]{Masucci2011Extracting}%
  \BibitemOpen
  \bibfield  {author} {\bibinfo {author} {\bibfnamefont {A.~P.}\ \bibnamefont
  {Masucci}}, \bibinfo {author} {\bibfnamefont {A.}~\bibnamefont {Kalampokis}},
  \bibinfo {author} {\bibfnamefont {V.~M.}\ \bibnamefont {Eguíluz}},\ and\
  \bibinfo {author} {\bibfnamefont {E.}~\bibnamefont {Hernández-García}},\
  }\bibfield  {title} {\bibinfo {title} {Extracting directed information flow
  networks: an application to genetics and semantics},\ }\href@noop {}
  {\bibfield  {journal} {\bibinfo  {journal} {Physical Review E}\ }\textbf
  {\bibinfo {volume} {83}},\ \bibinfo {pages} {026103} (\bibinfo {year}
  {2011})}\BibitemShut {NoStop}%
\bibitem [{\citenamefont {Gerlach}\ \emph {et~al.}(2015)\citenamefont
  {Gerlach}, \citenamefont {Fontclos},\ and\ \citenamefont
  {Altmann}}]{Gerlach2015Similarity}%
  \BibitemOpen
  \bibfield  {author} {\bibinfo {author} {\bibfnamefont {M.}~\bibnamefont
  {Gerlach}}, \bibinfo {author} {\bibfnamefont {F.}~\bibnamefont {Fontclos}},\
  and\ \bibinfo {author} {\bibfnamefont {E.~G.}\ \bibnamefont {Altmann}},\
  }\bibfield  {title} {\bibinfo {title} {On the similarity of symbol frequency
  distributions with heavy tails},\ }\href@noop {} {\bibfield  {journal}
  {\bibinfo  {journal} {Pakistan Journal of Pharmaceutical Sciences}\ }\textbf
  {\bibinfo {volume} {28}},\ \bibinfo {pages} {671} (\bibinfo {year}
  {2015})}\BibitemShut {NoStop}%
\bibitem [{\citenamefont {Everitt}\ \emph {et~al.}()\citenamefont {Everitt},
  \citenamefont {Skrondal},\ and\ \citenamefont {Booksx}}]{EverittThe}%
  \BibitemOpen
  \bibfield  {author} {\bibinfo {author} {\bibfnamefont {B.}~\bibnamefont
  {Everitt}}, \bibinfo {author} {\bibfnamefont {A.}~\bibnamefont {Skrondal}},\
  and\ \bibinfo {author} {\bibfnamefont {I.}~\bibnamefont {Booksx}},\
  }\bibfield  {title} {\bibinfo {title} {The cambridge dictionary of
  statistics, fourth edition},\ }\href@noop {} {\ }\BibitemShut {NoStop}%
\end{thebibliography}%

\end{document}